\documentstyle[11pt]{article}

\begin{document}

\title{{\Large {\bf Hypersurface homogeneous locally rotationally symmetric
spacetimes admitting conformal symmetries}}}
\author{Pantelis S. Apostolopoulos and Michael Tsamparlis \\
{\small {\it Department of Physics, Section of
Astrophysics-Astronomy-Mechanics},}\\
{\small {\it University of Athens, Panepistemiopolis, Athens 157 83, GREECE}}}
\maketitle

\begin{abstract}
All hypersurface homogeneous locally rotationally symmetric spacetimes which
admit conformal symmetries are determined and the symmetry vectors are given
explicitly. It is shown that these spacetimes must be considered in two
sets. One set containing Ellis Class II and the other containing Ellis Class
I, III LRS spacetimes. The determination of the conformal algebra in the
first set is achieved by systematizing and completing results on the
determination of CKVs in 2+2 decomposable spacetimes. In the second set new
methods are developed. The results are applied to obtain the classification
of the conformal algebra of {\em all} static LRS spacetimes in terms of
geometrical variables. Furthermore {\em all perfect fluid nontilted} LRS
spacetimes which admit proper conformal symmetries are determined and the
physical properties some of them are discussed.

{\bf PACS: 2.40.Hw, 4.20.-q, 4.20.Jb}
\end{abstract}

\section{Introduction}

\setcounter{equation}{0}

Hypersurface homogeneous spacetimes which are locally rotationally symmetric
(to be referred in the following as LRS spacetimes) contain many well known
and important families of exact solutions of Einstein field equations and
have been studied extensively in the literature \cite
{Ellis1,Stewart-Ellis,Ellis-MacCallum,MacCallum1,Collins1,Elst-EllisI}. They
admit a group of motions $G_{4}$ acting multiply transitively on three
dimensional orbits spacelike ($S_{3})$ or timelike ($T_{3})$ and the
isotropy group is a spatial rotation. It is well known that the metrics of
these spacetimes are \cite{Ellis1,Stewart-Ellis}:

\begin{equation}
ds^2=\varepsilon [dt^2-A^2(t)dx^2]+B^2(t)\left[ dy^2+\Sigma
^2(y,k)dz^2\right]  \label{sx1.1}
\end{equation}

\begin{equation}
ds^2=\varepsilon \left\{ dt^2-A^2(t)\left[ dx+\Lambda (y,k)dz\right]
^2\right\} +B^2(t)\left[ dy^2+\Sigma ^2(y,k)dz^2\right]  \label{sx1.2}
\end{equation}

\begin{equation}
ds^{2}=\varepsilon [dt^{2}-A^{2}(t)dx^{2}]+B^{2}(t)e^{2x}(dy^{2}+dz^{2})
\label{sx1.3}
\end{equation}
where $\varepsilon =\pm 1,\Sigma (y,k)=\sin y,\sinh y,y$ and $\Lambda
(y,k)=\cos y,\cosh y,y^{2}$ for $k=1,-1,0$ respectively. (The factor $%
\varepsilon =\pm 1$ essentially distinguishes between the ''static'' and the
''nonstatic'' cases as it can be seen by interchanging the co-ordinates $t,x$%
). According to the classification given by Ellis \cite{Ellis1} the metrics (%
\ref{sx1.2}) with $\varepsilon =1$ are Class $I$ LRS metrics, the metrics (%
\ref{sx1.1}) and (\ref{sx1.3}) are Class $II$ and finally Class $III$ are
the metrics (\ref{sx1.2}) with $\varepsilon =-1.$

The field equations for the LRS spacetimes reduce to a system of ordinary
differential equations which, in general, have not been integrated. Thus
additional simplifications have to be made and this is done by the
introduction of extra conditions which are constraints on the set of LRS
metrics (\ref{sx1.1})-(\ref{sx1.3}). The most important types of such
conditions have the form ${\cal L}_{{\bf \xi }}{\bf A}={\bf F}$ where ${\bf A%
}$ is any of the quantities $g_{ab},\Gamma _{bc}^{a},R_{ab},R_{bcd}^{a}$ and
geometric objects constructed by them and ${\bf F}$ is a tensor with the
same index symmetries as ${\bf A}$. These conditions are called {\em %
geometrical symmetries} or {\em collineations}. There are many types of
collineations and in fact most (but not all) of them have been classified in
an appropriate tree diagram \cite{Katzin-Levine-Davies} (corrected later in 
\cite{Katzin-Levine}).

One important and widely studied geometrical symmetry is the Conformal
Killing Vector (CKV). A CKV\ is defined by the requirement ${\cal L}_{{\bf %
\xi }}g_{ab}=2\psi ({\bf \xi })g_{ab}$ and specializes to a Killing Vector
(KV) ($\psi ({\bf \xi })=0$), to a Homothetic vector field (HVF) ($\psi (%
{\bf \xi })=$const.$\neq 0$) and to a Special Conformal Killing Vector
(SCKV) ($\psi _{;ab}=0$).

The purpose of the present paper is threefold:

a. To determine the metrics (\ref{sx1.1})-(\ref{sx1.3}) which admit (proper
or not) CKVs. We use/develop purely geometrical methods hence the results
hold for any type of matter.

b. To apply these results in two directions. First to classify in a natural
and geometric manner the CKVs of a static spherically symmetric spacetime 
\cite{MMT1,MMT2} and some other minor results \cite{Qadir-Ziad,Ahmad-Ziad}.
Secondly to find {\em all, physically acceptable, nontilted LRS perfect
fluid spacetimes which admit proper CKVs }and discuss briefly their basic
physical properties.

c. To find the CKVs which inherit the symmetry. An inheriting CKV $\xi ^{a}$
in a fluid spacetime with a four velocity $u^{a}$ is defined by the
requirement ${\cal L}_{{\bf \xi }}u^{a}=-\psi u^{a}$ where $\psi $ is the
conformal factor of $\xi ^{a}$. The inheritance property is important
because it assures that the Lie drag of the fluid flow lines by the CKV will
transform fluid flow lines onto fluid flow lines a fact that gives rise to
dynamical conservation laws and other useful kinematical and dynamical
results \cite
{Coley-Tupper1,Coley-Tupper2,Coley-Tupper3,Coley-Tupper4,Herrera1,Herrera2,Maartens-Mason-Tsamparlis1}%
.

The structure of the paper is as follows. In Section 2 we derive the CKVs of
the LRS metric (\ref{sx1.1}). We restrict our study to the nonstatic case ($%
\varepsilon =1$), because the results of the static case ($\varepsilon =-1$)
can be obtained by interchanges the co-ordinates $t,x$. In Section 3 we
apply this approach to classify the conformal algebra of a static
spherically symmetric spacetime. In Sections 4 and 5 we derive the complete
conformal algebra of the remaining metrics of Ellis class II and the Ellis
class I,III. In Section 6 we determine all LRS\ perfect fluid spacetimes
which admit a proper CKV and satisfy the weak and the dominant energy
conditions. Finally Section 7 concludes the paper.

\section{The conformal algebra of the LRS metrics (1.1)}

\setcounter{equation}{0}

The LRS spacetimes with metric (\ref{sx1.1}) and $\varepsilon =-1$ admit the
isometry group $G_{4}$ consisting of the four KVs $\partial _{x},{\bf X}%
_{\mu }$ $(\mu =1,2,3)$: 
\begin{equation}
{\bf X}_{\mu }=(\delta _{\mu }^{1}\cos z+\delta _{\mu }^{2}\sin z)\partial
_{y}-\left[ (\ln \Sigma )_{,y}(\delta _{\mu }^{1}\sin z-\delta _{\mu
}^{2}\cos z)-\delta _{\mu }^{3}\right] \partial _{z}  \label{sx2.1}
\end{equation}
acting on 3D spacelike orbits. The metric (\ref{sx1.1}) can be written:

\begin{equation}
ds^{2}=B^{2}(t)d\bar{s}^{2}  \label{sx2.2}
\end{equation}
where:

\begin{equation}
d\bar{s}^{2}=-\frac{dt^{2}}{B^{2}(t)}+\frac{A^{2}(t)}{B^{2}(t)}%
dx^{2}+dy^{2}+\Sigma ^{2}(y,k)dz^{2}  \label{sx2.3}
\end{equation}
is a \{2+2\} decomposable metric whose constituent 2-spaces are:

\begin{equation}
d\Omega ^{2}=dy^{2}+\Sigma ^{2}(y,k)dz^{2}  \label{sx2.4}
\end{equation}
with constant curvature $R_{2}=2k$ ($k=0,1,-1$) and:

\begin{equation}
ds_{L}^{2}=-\frac{dt^{2}}{B^{2}(t)}+\frac{A^{2}(t)}{B^{2}(t)}dx^{2}
\label{sx2.5}
\end{equation}
with scalar curvature $R_{1}$ given by:

\begin{equation}
R_{1}=2\left( B\frac{dq_{1}}{dt}+q_{1}^{2}\right)  \label{sx2.6}
\end{equation}
where:

\begin{equation}
q_{1}=B\frac{d\left[ \ln \left( \frac{A}{B}\right) \right] }{dt}.
\label{sx2.7}
\end{equation}
From (\ref{sx2.2}) it follows that in order to determine the conformal
algebra of the LRS metrics (\ref{sx1.1}) it suffices to determine the
conformal algebra of the \{2+2\} metric (\ref{sx2.3}) (we recall that the
conformal factors $\psi ^{\prime }({\bf X}),\psi ({\bf X})$ of a CKV ${\bf X}
$ of two conformally related metrics $g_{ij}^{\prime }=N^{2}g_{ij}$ satisfy
the equation $\psi ^{\prime }({\bf X})=\psi ({\bf X})+{\bf X}(\ln N)$.)

Regarding the determination of the KVs and HVFs of a \{2+2\} decomposable
metric there are standard results in the literature. In fact it has been
shown that the KVs of each of the constituent 2-metrics are KVs of the whole
\{2+2\} spacetime. Furthermore a \{2+2\} spacetime admits a HVF with
homothetic factor $bb^{\prime }$ if and only if each 2-metric admits one
with corresponding homothetic factor $b,b^{\prime }$ \cite{Hall-daCosta}.

Concerning the proper CKVs Coley and Tupper \cite{Coley-Tupper5} have proved
that \{2+2\} decomposable spacetimes admit proper CKVs if (a) their
constituent 2-spaces are spaces of constant {\em nonvanishing} curvature, $%
R_{1}$ and $R_{2}$ say, such that $R_{1}+R_{2}=0$ and/or (b) they are
conformally flat (these spacetimes are the Bertotti-Robinson and
''anti-Bertotti-Robinson'' spacetimes). However Coley and Tupper did not
give a method to determine the CKVs explicitly. The following Proposition 1
provides such a method by showing that a {\em proper} CKV in a \{2+2\}
spacetime is expressed in terms of the {\em gradient} CKVs of the
constituent 2-metrics.

{\bf Proposition 1} {\em Let }$g_{ab}=g_{AB}\otimes g_{A^{\prime }B^{\prime
}}${\em \ be a \{2+2\} decomposable spacetime where }$A,...=0,1,$ $A^{\prime
},...=2,3,$ $a,b=0,1,2,3${\em \ and }$g_{AB}(x^{C})${\em \ }$(${\em \
respectively }$g_{A^{\prime }B^{\prime }}(x^{C^{\prime }}))${\em \ is the
Lorentzian (respectively Euclidean) part. If both 2-metrics are metrics of
constant but opposite nonvanishing curvature i.e. }$R_{1}=-R_{2}=2p${\em \ (}%
$p\neq 0${\em ) then the \{2+2\} metric }$g_{ab}${\em \ is conformally flat
and admits 15 CKVs. The conformal algebra contains the six KVs (three KVs
for each 2-metric) plus the nine proper CKVs }${\bf \xi }_{\alpha \beta }$%
{\em \ given by (}$\alpha ,\beta =1,2,3${\em ): } 
\begin{equation}
{\bf \xi }_{\alpha \beta }=-\frac{1}{p}\left[ f_{\alpha }^{\prime }(f_{\beta
})^{,A}\partial _{A}-f_{\beta }\left( f_{\alpha }^{\prime }\right)
^{,A^{\prime }}\partial _{A^{\prime }}\right]  \label{sx2.7a}
\end{equation}
{\em with conformal factor }$\psi ({\bf \xi }_{\alpha \beta })=f_{\alpha
}^{\prime }f_{\beta }$ ,{\em where }$f(x^{A}),f^{\prime }(x^{A^{\prime }})$%
{\em \ are smooth (independent) and real valued functions such that: }

\begin{equation}
f_{|AB}=-pfg_{AB}\mbox{\qquad and\qquad }f_{||A^{\prime }B^{\prime
}}^{\prime }=pf^{\prime }g_{A^{\prime }B^{\prime }}  \label{sx2.7b}
\end{equation}
{\em and ''}$|${\em '', ''}$||${\em '' denote covariant differentiation with
respect to the metrics }$g_{AB},g_{A^{\prime }B^{\prime }}${\em \
respectively. }

From the above discussion becomes clear that one has to consider two cases:

Case A Metrics (\ref{sx2.3}) which are not conformally flat and possibly
admit KVs and HVFs only.

Case B The conformally flat Bertotti-Robinson and ''anti-Bertotti-Robinson''
spacetimes and, of course, the Minkowski spacetime.

In each case the exact form of the resulting \{2+2\} spacetime is determined
by solving the ordinary differential equation (\ref{sx2.6}).

{\bf Case A}

In this case the reduced \{2+2\} metric (\ref{sx2.3}) does not admit proper
CKVs and there are only two possibilities to consider:

a) The \{2+2\} metric admits a HVF. This is possible if and only if both
2-metrics admit a HVF \cite{Hall-daCosta}. This occurs only if $k=0$ and the
metric (\ref{sx2.5}) is not of constant curvature. Furthermore because the
Lie bracket of an HVF with a KV is a KV, using Jacobi identities and the
original symmetry equations, we can determine the form of the resulting
\{2+2\} metric and the HVF.

b) The \{2+2\} metric admits two extra KVs acting on 2D timelike orbits
hence the 2-space (\ref{sx2.5}) becomes a space of constant curvature $%
R_1\neq -R_2$. Setting $R_1=2\epsilon /a^2$ where $\epsilon =0,\pm 1$ and
using (\ref{sx2.6}) we determine the 2D metric (\ref{sx2.5}) whose isometry
algebra is computed in a straightforward manner.

The rather lengthy and typical calculations present no particular interest
and we summarize the results in Table 1 where it can be seen that there are
seven different cases $A_{1},...,A_{7}.$ \pagebreak

{\bf Table 1. }Proper CKVs admitted by the nonconformally flat LRS metrics (%
\ref{sx1.1}). In all cases $\tau =\int \frac{dt}{B(t)}$ and $R_{1}$ is the
curvature of the 2-space (\ref{sx2.5}). We note that in cases A$_{3}$,A$_{4}$
and A$_{5}$ the curvature $R_{1}\neq 2$ whereas in case A$_{6}$ the
curvature $R_{1}\neq -2$. Furthermore the vectors ${\bf \xi ,\xi }_{2}$ of
cases A$_{1}$ and A$_{2}$,A$_{3}$ respectively are inheriting, $\mu =2,3$
and $\varepsilon _{1}=\pm 1$, $\alpha _{1}\neq 0,1$ in order to avoid the
conformal flatness of the metric.

\begin{center}
\begin{tabular}{|c|c|c|c|c|c|}
\hline
{\bf Case} & $\frac{A(\tau )}{B{\bf (}\tau {\bf )}}$ & $R_{1}$ & $k$ & {\bf %
CKVs} & {\bf Conformal Factor} \\ \hline
A$_{1}$ & $\tau ^{\frac{\alpha _{1}-1}{\alpha _{1}}}$ & $-$ & $0$ & ${\bf %
\xi }=\alpha _{1}\tau \partial _{\tau }+x\partial _{x}+\alpha _{1}y\partial
_{y}$ & $\alpha _{1}\left[ 1+\tau (\ln B)_{,\tau }\right] $ \\ \hline
A$_{2}$ & $1$ & $0$ & $\pm 1$ & ${\bf \xi }_{\mu }=\delta _{\mu
}^{2}\partial _{\tau }+\delta _{\mu }^{3}(x\partial _{\tau }+\tau \partial
_{x})$ & $\left[ \delta _{\mu }^{2}+\delta _{\mu }^{3}x\right] (\ln
B)_{,\tau }$ \\ \hline
A$_{3}$ & $e^{\varepsilon _{1}\tau /a}$ & $\frac{2}{a^{2}}$ & $0,\pm 1$ & $
\begin{array}{l}
{\bf \xi }_{\mu }=\delta _{\mu }^{2}(-\varepsilon _{1}a\partial _{\tau
}+x\partial _{x})+ \\ 
+\delta _{\mu }^{3}\{-2\varepsilon _{1}ax\partial _{\tau }+\left[
x^{2}+a^{2}e^{-2\varepsilon _{1}\tau /a}\right] \partial _{x}\}
\end{array}
$ & $
\begin{array}{l}
-\delta _{\mu }^{2}\varepsilon _{1}a(\ln B)_{,\tau }- \\ 
-\delta _{\mu }^{3}2x\varepsilon _{1}a(\ln B)_{,\tau }
\end{array}
$ \\ \hline
A$_{4}$ & $\cosh \frac{\tau }{a}$ & $\frac{2}{a^{2}}$ & $0,\pm 1$ & $
\begin{array}{c}
{\bf \xi }_{\mu }=\delta _{\mu }^{2}\left[ \sin (\frac{x}{a})\partial _{\tau
}+\tanh (\frac{\tau }{a})\cos (\frac{x}{a})\partial _{x}\right] + \\ 
+\delta _{\mu }^{3}\left[ \cos (\frac{x}{a})\partial _{\tau }-\tanh (\frac{%
\tau }{a})\sin (\frac{x}{a})\partial _{x}\right]
\end{array}
$ & $
\begin{array}{c}
\delta _{\mu }^{2}\sin (\frac{x}{a})(\ln B)_{,\tau }+ \\ 
\delta _{\mu }^{3}\cos (\frac{x}{a})(\ln B)_{,\tau }
\end{array}
$ \\ \hline
A$_{5}$ & $\sinh \frac{\tau }{a}$ & $\frac{2}{a^{2}}$ & $0,\pm 1$ & $
\begin{array}{c}
{\bf \xi }_{\mu }=\delta _{\mu }^{2}\left[ \sinh (\frac{x}{a})\partial
_{\tau }-\coth (\frac{\tau }{a})\cosh (\frac{x}{a})\partial _{x}\right] + \\ 
+\delta _{\mu }^{3}\left[ \cosh (\frac{x}{a})\partial _{\tau }-\coth (\frac{%
\tau }{a})\sinh (\frac{x}{a})\partial _{x}\right]
\end{array}
$ & $
\begin{array}{c}
\delta _{\mu }^{2}\sinh (\frac{x}{a})(\ln B)_{,\tau }+ \\ 
\delta _{\mu }^{3}\cosh (\frac{x}{a})(\ln B)_{,\tau }
\end{array}
$ \\ \hline
A$_{6}$ & $\cos \frac{\tau }{a}$ & $-\frac{2}{a^{2}}$ & $0,\pm 1$ & $
\begin{array}{c}
{\bf \xi }_{\mu }=\delta _{\mu }^{2}\left[ \sinh (\frac{x}{a})\partial
_{\tau }+\tan (\frac{\tau }{a})\cosh (\frac{x}{a})\partial _{x}\right] + \\ 
+\delta _{\mu }^{3}\left[ \cosh (\frac{x}{a})\partial _{\tau }+\tan (\frac{%
\tau }{a})\sinh (\frac{x}{a})\partial _{x}\right]
\end{array}
$ & $
\begin{array}{c}
\delta _{\mu }^{2}\sinh (\frac{x}{a})(\ln B)_{,\tau }+ \\ 
\delta _{\mu }^{3}\cosh (\frac{x}{a})(\ln B)_{,\tau }
\end{array}
$ \\ \hline
A$_{7}$ & $\tau $ & $0$ & $\pm 1$ & $
\begin{array}{c}
{\bf \xi }_{\mu }=\delta _{\mu }^{2}\left[ \cosh x\partial _{\tau }-\tau
^{-1}\sinh x\partial _{x}\right] + \\ 
+\delta _{\mu }^{3}\left[ \sinh x\partial _{\tau }-\tau ^{-1}\cosh x\partial
_{x}\right]
\end{array}
$ & $
\begin{array}{c}
\delta _{\mu }^{2}\cosh x(\ln B)_{,\tau }+ \\ 
\delta _{\mu }^{3}\sinh x(\ln B)_{,\tau }
\end{array}
$ \\ \hline
\end{tabular}
\end{center}

{\bf Case B}

The conformal algebra of the Bertotti-Robinson and the
''anti-Bertotti-Robinson'' spacetime consists of 6 KVs (three KVs for each
2-space of constant curvature) plus nine {\em proper} CKVs and it is
isomorphic to the conformal algebra SO(4,2) of Minkowski spacetime. The
explicit computation of the CKVs in the coordinates in which the above
metrics are defined can be done easily by means of Proposition 1.

The Minkowski spacetime results in two different types of LRS spacetimes
that is $A(\tau )=B(\tau )$ and $A(\tau )=\tau B(\tau )$. In the former case
the resulting metric is the flat Robertson-Walker metric whose conformal
algebra is known \cite{Maa-Mah}. The second LRS metric is:

\begin{equation}
ds^{2}=B^{2}(\tau )\left( -d\tau ^{2}+\tau
^{2}dx^{2}+dy^{2}+y^{2}dz^{2}\right) .  \label{sx2.8}
\end{equation}
which by means of the transformation:

\[
\hat{t}=\tau \cosh x,\qquad \hat{x}=\tau \sinh x,\qquad \hat{y}=y\cos
z,\qquad \hat{z}=y\sin z 
\]
reduces to the metric $B(\hat{t},\hat{x})(-d\hat{t}^{2}+d\hat{x}^{2}+d\hat{y}%
^{2}+d\hat{z}^{2})$ whose conformal algebra is also known \cite{Bruhat}.

In Table 2 we collect all conformally flat LRS metrics and their
corresponding CKVs (except the trivial case of Minkowski
spacetime).\pagebreak

{\bf Table 2}. Proper CKVs of the conformally flat LRS metrics (\ref{sx1.1}%
). In case B$_{1}$ the spacetime is spherically symmetric ($k=1$) and the
functions $f_{\alpha }^{\prime }=\left( -\cos y,\sin y\cos z,\sin y\sin
z\right) $. The other cases correspond to hyperbolic spacetimes ($k=-1$) and
the functions $f_{\alpha }^{\prime }=\left( \cosh y,\sinh y\cos z,\sinh
y\sin z\right) $. The nontensorial indices $\alpha ,\mu =1,2,3$ and $A=\tau
,x$ and $A^{\prime }=y,z$.

\begin{center}
\small{ 
\begin{tabular}{|c|c|c|c|c|}
\hline
{\bf Case} & $\frac{{\bf A(\tau )}}{{\bf B(\tau )}}$ & {\bf Conformal
Killing Vectors} & $f$ & {\bf Conformal Factors} \\ \hline
B$_1$ & $\cos \tau $ & $
\begin{array}{c}
{\bf X}_2=\sinh x\partial _\tau +\tan \tau \cosh x\partial _x \\ 
{\bf X}_3=\cosh x\partial _\tau +\tan \tau \sinh x\partial _x \\ 
\begin{tabular}{l}
${\bf X}_{3(\alpha +1)+\mu }$ \\ 
$=f_\mu ^{\prime }\left( f_\alpha \right) ^{,A}\partial _A-f_\alpha \left(
f_\mu ^{^{\prime }}\right) ^{,A^{\prime }}\partial _{A^{\prime }}$%
\end{tabular}
\end{array}
$ & $
\begin{tabular}{l}
$f_1=-\sin \tau $ \\ 
$f_2=\cos \tau \cosh x$ \\ 
$f_3=\cos \tau \sinh x$%
\end{tabular}
$ & $
\begin{array}{c}
\psi ({\bf X}_2)=\sinh x(\ln B)_{,\tau } \\ 
\psi ({\bf X}_3)=\cosh x(\ln B)_{,\tau } \\ 
\begin{tabular}{l}
$\psi ({\bf X}_{3(\alpha +1)+\mu })=f_\mu ^{\prime }f_\alpha $ \\ 
$\times \left[ -(\ln \left| f_\alpha \right| )_{,\tau }(\ln B)_{,\tau
}+1\right] $%
\end{tabular}
\end{array}
$ \\ \hline
B$_2$ & $e^{\varepsilon _1\tau }$ & $
\begin{array}{c}
{\bf X}_2=-\varepsilon _1\partial _\tau +x\partial _x \\ 
{\bf X}_3=-2\varepsilon _1x\partial _\tau +\left[ x^2+e^{-2\varepsilon
_1\tau }\right] \partial _x \\ 
\begin{tabular}{l}
${\bf X}_{3(\alpha +1)+\mu }$ \\ 
$=-\left[ f_\mu ^{\prime }\left( f_\alpha \right) ^{,A}\partial _A-f_\alpha
\left( f_\mu ^{^{\prime }}\right) ^{,A^{\prime }}\partial _{A^{\prime
}}\right] $%
\end{tabular}
\end{array}
$ & $
\begin{tabular}{l}
$f_1=-e^{\varepsilon _1\tau }$ \\ 
$f_2=-e^{\varepsilon _1\tau }x$ \\ 
$f_3=-e^{\varepsilon _1\tau }(x^2-e^{-2\varepsilon _1\tau })$%
\end{tabular}
$ & $
\begin{array}{c}
\psi ({\bf X}_2)=-\varepsilon _1(\ln B)_{,\tau } \\ 
\psi ({\bf X}_3)=-2x\varepsilon _1(\ln B)_{,\tau } \\ 
\begin{tabular}{l}
$\psi ({\bf X}_{3(\alpha +1)+\mu })=-f_\mu ^{\prime }f_\alpha $ \\ 
$\times \left[ -(\ln \left| f_\alpha \right| )_{,\tau }(\ln B)_{,\tau
}-1\right] $%
\end{tabular}
\end{array}
$ \\ \hline
B$_3$ & $\cosh \tau $ & $
\begin{array}{c}
{\bf X}_2=\sin x\partial _\tau +\tanh \tau \cos x\partial _x \\ 
{\bf X}_3=\cos x\partial _\tau +\tanh \tau \sin x\partial _x \\ 
\begin{tabular}{l}
${\bf X}_{3(\alpha +1)+\mu }$ \\ 
$=-\left[ f_\mu ^{\prime }\left( f_\alpha \right) ^{,A}\partial _A-f_\alpha
\left( f_\mu ^{^{\prime }}\right) ^{,A^{\prime }}\partial _{A^{\prime
}}\right] $%
\end{tabular}
\end{array}
$ & $
\begin{tabular}{l}
$f_1=-\sinh \tau $ \\ 
$f_2=-\cosh \tau \cos x$ \\ 
$f_3=-\cosh \tau \sin x$%
\end{tabular}
$ & $
\begin{array}{c}
\psi ({\bf X}_2)=\sin x(\ln B)_{,\tau } \\ 
\psi ({\bf X}_3)=\cos x(\ln B)_{,\tau } \\ 
\begin{tabular}{l}
$\psi ({\bf X}_{3(\alpha +1)+\mu })=-f_\mu ^{\prime }f_\alpha $ \\ 
$\times \left[ -(\ln \left| f_\alpha \right| )_{,\tau }(\ln B)_{,\tau
}-1\right] $%
\end{tabular}
\end{array}
$ \\ \hline
B$_4$ & $\sinh \tau $ & $
\begin{array}{c}
{\bf X}_2=\sinh x\partial _\tau -\coth \tau \cosh x\partial _x \\ 
{\bf X}_3=\cosh x\partial _\tau -\coth \tau \sinh x\partial _x \\ 
\begin{tabular}{l}
${\bf X}_{3(\alpha +1)+\mu }$ \\ 
$=-\left[ f_\mu ^{\prime }\left( f_\alpha \right) ^{,A}\partial _A-f_\alpha
\left( f_\mu ^{^{\prime }}\right) ^{,A^{\prime }}\partial _{A^{\prime
}}\right] $%
\end{tabular}
\end{array}
$ & $
\begin{tabular}{l}
$f_1=-\cosh \tau $ \\ 
$f_2=-\sinh \tau \cosh x$ \\ 
$f_3=-\sinh \tau \sinh x$%
\end{tabular}
$ & $
\begin{array}{c}
\psi ({\bf X}_2)=\sinh x(\ln B)_{,\tau } \\ 
\psi ({\bf X}_3)=\cosh x(\ln B)_{,\tau } \\ 
\begin{tabular}{l}
$\psi ({\bf X}_{3(\alpha +1)+\mu })=-f_\mu ^{\prime }f_\alpha $ \\ 
$\times \left[ -(\ln \left| f_\alpha \right| )_{,\tau }(\ln B)_{,\tau
}-1\right] $%
\end{tabular}
\end{array}
$ \\ \hline
\end{tabular}
}
\end{center}
\pagebreak 

\subsection{Discussion of the results on the LRS metrics (1.1)}

{\em 1. Homogeneous LRS spacetimes}

From Table 1 it follows that the nonconformally flat LRS spacetimes (\ref
{sx1.1}) admit either one or two extra KVs. It is well known that the only
perfect fluid LRS spacetime with a maximal $G_{5}$ is the G\"{o}del solution
which is of Ellis class I. Furthermore $\Lambda $-term solutions with $G_{5}$
as a maximal group of motions are of Petrov type N \cite{Kramer}. Therefore
the homogeneous LRS metric A$_{1}$ of Table\ 1 can represent anisotropic
fluid spacetimes only. Regarding the remaining nonconformally flat
homogeneous spacetimes of Table 1 they can represent either a $\Lambda $%
-term solution or anisotropic fluid solutions.

From Table\ 2 we see that the conformally flat homogeneous LRS metrics
consist of three different types:

a. Metrics of constant (positive) curvature which admit 10 KVs (de Sitter
spacetime)

b. Metrics which are not of constant curvature and admit 7 KVs and

c. The Bertotti-Robinson and ''anti-Bertotti-Robinson'' spacetimes which
admit a $G_{6}$ group of motions.

LRS metrics of type a and c are well known whereas type b LRS metrics have
been found by Rebou\c{c}as and Tiomno (the RT spacetime) \cite{RT} and Rebou%
\c{c}as and Texeira (the ''anti RT'' or ART spacetime) \cite{ART}. They are
1+3 decomposable spacetimes the 3-spaces of which are timelike spaces of
constant curvature (negative for RT and positive for ART). Their analogue is
the Einstein and the ''anti-Einstein'' spacetimes whose 3-spaces are
spacelike surfaces of constant curvature. All homogeneous LRS metrics are
given in Table 3. We should not that ''anti-Einstein'', ART and
''anti-Bertotti-Robinson'' metrics {\em do not satisfy} the energy
conditions. \bigskip

{\bf Table 3. }This Table contains all homogeneous LRS spacetimes with
metric (1.1). The indices $\alpha ,\mu =2,3$ and the constants $%
a,c,\varepsilon _{1}$ satisfy the constraints $ac\neq 0$ and $\varepsilon
_{1}=\pm 1$.

\begin{tabular}{|l|l|l|l|l|l|}
\hline
{\bf Case} & $k$ & $A(t)$ & $B(t)$ & {\bf KVs} & {\bf Type of the metric} \\ 
\hline
A$_1$ & $0$ & $ce^{-t/c\alpha _1}$ & $ce^{-t/c}$ & ${\bf \xi }$ & LRS \\ 
\hline
A$_2$ & $\pm 1$ & $c_1c_2$ & $c_2$ & ${\bf \xi }_\mu $ & 1+1+2 \\ \hline
A$_3$ & $0,\pm 1$ & $ce^{\varepsilon _1t/ac}$ & $c$ & ${\bf \xi }_\mu $ & 2+2
\\ \hline
A$_4$ & $0,\pm 1$ & $c\cosh \frac t{ca}$ & $c$ & ${\bf \xi }_\mu $ & 2+2 \\ 
\hline
A$_5$ & $0,\pm 1$ & $c\sinh \frac t{ca}$ & $c$ & ${\bf \xi }_\mu $ & 2+2 \\ 
\hline
A$_6$ & $0,\pm 1$ & $c\cos \frac t{ca}$ & $c$ & ${\bf \xi }_\mu $ & 2+2 \\ 
\hline
A$_7$ & $\pm 1$ & $ct$ & $c$ & ${\bf \xi }_\mu $ & 1+1+2 \\ \hline
B$_1$ & $1$ & $c\cot \tau $ & $\frac c{\sin \tau }$ & ${\bf X}_{3(\alpha
+1)+\mu }$ & Constant Curvature (Type a) \\ \hline
B$_3$ & $-1$ & $c\coth \tau $ & $\frac c{\sinh \tau }$ & ${\bf X}_{3(\alpha
+1)+\mu }$ & Constant Curvature (Type a) \\ \hline
B$_4$ & $-1$ & $c\tanh \tau $ & $\frac c{\cosh \tau }$ & ${\bf X}_{3(\alpha
+1)+\mu }$ & Constant Curvature (Type a) \\ \hline
B$_1$ & $1$ & $c$ & $\frac c{\cos \tau }$ & ${\bf X}_{6+\mu }$ & 1+3 (Type b)
\\ \hline
B$_3$ & $-1$ & $c$ & $\frac c{\cosh \tau }$ & ${\bf X}_{6+\mu }$ & 1+3 (Type
b) \\ \hline
B$_4$ & $-1$ & $c$ & $\frac c{\sinh \tau }$ & ${\bf X}_{6+\mu }$ & 1+3 (Type
b) \\ \hline
B$_1$ & $1$ & $c\cos \frac tc$ & $c$ & ${\bf X}_\mu $ & 2+2 (Type c) \\ 
\hline
B$_2$ & $-1$ & $ce^{\varepsilon _1t/c}$ & $c$ & ${\bf X}_\mu $ & 2+2 (Type c)
\\ \hline
B$_3$ & $-1$ & $c\cosh \frac tc$ & $c$ & ${\bf X}_\mu $ & 2+2 (Type c) \\ 
\hline
B$_4$ & $-1$ & $c\sinh \frac tc$ & $c$ & ${\bf X}_\mu $ & 2+2 (Type c) \\ 
\hline
\end{tabular}
\bigskip

{\em 2. LRS spacetimes admitting a transitive homothety group }$H_{5}$

These spacetimes follow directly from the metrics of Table 1 if we set the
conformal factor to be a constant. These metrics have been found by Carot
and Sintes \cite{Carot-Sintes} and Sintes \cite{Sintes} and need not be
considered further (we should note that the HVF ${\bf \xi }$ of Case A$_{1}$
has also been found previously by Tupper \cite{Tupper1}). For completeness
we collect the resulting metrics together with their HVFs in Table\
4.
\pagebreak

{\bf Table 4. }LRS spacetimes (\ref{sx1.1}) with transitive homothety group $%
H_{5}$ and $\varepsilon _{1}=\pm 1$.

\begin{tabular}{|l|l|l|l|l|l|}
\hline
{\bf Case} & $k$ & $A(t)$ & $B(t)$ & {\bf HKVs} & {\bf Conformal Factor} \\ 
\hline
A$_1$ & $0$ & $t^{\frac{b-1}b}$ & $c_1t^{\frac{b-\alpha _1}b}$ & ${\bf \xi }%
=bt\partial _t+x\partial _x+\alpha _1y\partial _y$ & $b$ \\ \hline
A$_2$ & $\pm 1$ & $\alpha _1t$ & $\alpha _1t$ & ${\bf \xi }=\alpha
_1t\partial _t$ & $\alpha _1$ \\ \hline
A$_3$ & $0,\pm 1$ & $(\alpha _1t)^{(1+\frac{\varepsilon _1}{c_1a})}$ & $%
\alpha _1t$ & ${\bf \xi }=-\varepsilon _1a\alpha _1t\partial _t+x\partial _x$
& $-\varepsilon _1a\alpha _1$ \\ \hline
\end{tabular}
\bigskip

{\em 3. Comparison with existing results}

Although many nonstatic LRS metrics of type (\ref{sx1.1}) are known which
are homogeneous or admit a HVF, it appears that there do not exist many such
metrics which admit proper CKVs. The metric found by Maartens and Mellin 
\cite{Maa-Mellin} belongs to the case A$_{6}$ ($k=0$) and their CKV equals $%
{\bf \xi }_{2}+{\bf \xi }_{3}$ with conformal factor $\psi ({\bf \xi }_{,2}+%
{\bf \xi }_{,3})=e^{x/a}(\ln B)_{,\tau }=e^{x/a}\dot{B}$ (the constant $a$
appearing in \cite{Maa-Mellin} must be replaced with $1/a$). Kitamura \cite
{Kitamura2} using a different method has found the conformal algebra of the
metrics (\ref{sx1.1}) for $k=1$ in the nonconformally flat case. Finally the
CKVs of the RT and ART spacetimes have been given in \cite
{Tsa-Nik-Apos,Tsa-Apos}.

\section{The static spherically symmetric case}

\setcounter{equation}{0}

One useful application of the results of the last section is the
classification of the conformal algebra of the static
spherically/plane/hyperbolically symmetric spacetimes. This is done if one
interchanges the co-ordinates $t,x$ and takes $k=1,0,-1$ respectively.

The classification of the proper CKVs of the static spherically symmetric
(SSS) spacetimes has been derived by Maartens et al. \cite{MMT1,MMT2} using
the direct (and more difficult) method of solving the differential conformal
equations ${\cal L}_{\xi }g_{ab}=2\psi g_{ab}$ in a SSS spacetime. We shall
show that the approach developed in the last section is straightforward,
more geometrical and results in a new inherent classification whose
classifying parameters have a direct geometrical interpretation.

The line element of a SSS\ spacetime has the general form \cite{Kramer}:

\begin{equation}
ds^2=-e^{2\nu (r)}dt^2+e^{2\lambda (r)}dr^2+r^2(d\theta ^2+\sin ^2\theta
d\phi ^2)  \label{sx3.1}
\end{equation}
and corresponds to the LRS\ metric (\ref{sx1.1}) with $\varepsilon =1,k=1$.
The analysis of the previous section applies provided one interchanges the
co-ordinates $t,r$. To show this we rewrite (\ref{sx3.1}) as follows:

\[
ds^2=r^2\left[ ds_L^2+(d\theta ^2+\sin ^2\theta d\phi ^2)\right] , 
\]
where $ds_L^2=\frac 1{r^2}(-e^{2\nu (r)}dt^2+e^{2\lambda (r)}dr^2)$. The
curvature $R_2$ of the 2-metric $ds_L^2$ is: 
\begin{equation}
R_2=-2(re^{-\lambda }q_2^{\prime }+q_2^2)  \label{sx3.2}
\end{equation}
where 
\begin{equation}
q_2=e^{-\lambda }(r\nu ^{\prime }-1)  \label{sx3.3}
\end{equation}
and a prime denotes differentiation w.r.t. the co-ordinate $r$.

The quantities $R_{2}$ and $q_{2}$ in this case are the parameters (\ref
{sx2.5}) and (\ref{sx2.6}) used for the computation and the classification
of the conformal algebra in the last Section 2. Therefore we can write the
form of the metric and the corresponding CKVs, using the results of Tables
1,2, provided that one makes the following correspondence:

\[
\bar{r}\leftrightarrow t,t\leftrightarrow \overline{r}\mbox{ where }%
\overline{r}=\int e^{\lambda }dr,A(r)=e^{\nu (r)},B(r)=\int e^{-\lambda }d%
\overline{r}\mbox{ and }\tau \leftrightarrow \widetilde{r}=\int \frac{d%
\overline{r}}{B(\overline{r})}. 
\]
For example in case A$_{6}$ the proper CKVs of the SSS spacetime are:

\[
\begin{array}{c}
{\bf \xi }_{2}=\sinh (\frac{t}{a})\partial _{\widetilde{r}}+\tan (\frac{%
\tilde{r}}{a})\cosh (\frac{t}{a})\partial _{t} \\ 
{\bf \xi }_{3}=\cosh (\frac{t}{a})\partial _{\widetilde{r}}+\tan (\frac{%
\tilde{r}}{a})\sinh (\frac{t}{a})\partial _{t}.
\end{array}
\]
\label{first}Therefore the method of Section 2 results in a complete
classification of the conformal algebra of SSS spacetimes in terms of the
parameters $R_{2},q_{2}$ which have a direct geometrical meaning.

This classification coincides with the scheme of Maartens et al. \cite{MMT1}
because the quantity $q$ used in \cite{MMT1} is related to $R_2$ by $%
q=-(1+\frac 12R_2)$ and the other classifying quantity $e^{-\lambda }(r\nu
^{\prime }-1)$ equals $q_2.$ Finally their constant $w=1/a^2$.

The CKVs found in \cite{MMT1} are linear combinations of the ones found in
this work. However, in some cases there is a direct correspondence. For
example the CKVs corresponding to the case $I_q[0]\leftrightarrow $A$_3$ are
given by ${\bf \xi }_2$ and ${\bf \xi }_3$.

In addition to the above, other results on the conformal algebra of SSS
spacetimes follow immediately form the present approach. For example the HVF
found by Ahmad and Ziad \cite{Ahmad-Ziad} is the vector ${\bf \xi }%
_{3}=-\varepsilon _{1}a\alpha _{1}t\partial _{t}+x\partial _{x}$ of the case
A$_{3}$ and the SSS spacetimes which admit six KVs \cite{Qadir-Ziad} follow
from those of Table 3 if we set the unimportant constant $A$ of \cite
{Qadir-Ziad} equal to zero.

\section{The CKVs of the LRS metrics (1.3)}

\setcounter{equation}{0}

The remaining LRS metrics of Ellis class II given in (\ref{sx1.3}) can be
written:

\begin{equation}
ds^{2}=e^{2x}B^{2}(t)ds_{2+2}^{2}  \label{sx4.1}
\end{equation}
where:

\begin{equation}
ds_{2+2}^{2}=e^{-2x}L^{2}(\bar{t})(-d\bar{t}^{2}+dx^{2})+dy^{2}+dz^{2}
\label{sx4.2}
\end{equation}
and $L(\bar{t})=\frac{A(\bar{t})}{B(\bar{t})}$ ($\bar{t}=\int \frac{dt}{A(t)}
$). Obviously the scenario of Section 2 applies again and one has to
consider the conformally flat and the nonconformally flat cases of the
reduced 2+2 spacetime.

{\bf Case A}

In this case we find that the metric (\ref{sx4.1}) admits the CKVs (which
are KV for the reduced 2+2 spacetime (\ref{sx4.2})):

\begin{equation}
{\bf \xi }=e^{-x}\left( \sinh \bar{t}\partial _{\bar{t}}-\cosh \bar{t}%
\partial _x\right)  \label{sx4.3}
\end{equation}
with $L{\bf (}\overline{t}{\bf )=}\sinh ^{-2}\bar{t}$ and conformal factor $%
\psi ({\bf \xi })=e^{-x}\left[ \sinh \bar{t}(\ln B)_{,\bar{t}}-\cosh \bar{t}%
\right] $ or,

\begin{equation}
{\bf \xi }=e^{-x}\left( \cosh \bar{t}\partial _{\bar{t}}-\sinh \bar{t}%
\partial _x\right)  \label{sx4.3a}
\end{equation}
with $L(\overline{t})=\cosh ^{-2}\bar{t}$ and conformal factor $\psi ({\bf %
\xi })=e^{-x}\left[ \cosh \bar{t}(\ln B)_{,\bar{t}}-\sinh \bar{t}\right] $.
Furthermore no inheriting proper CKV exists.

These CKVs reduce to KVs for $B(t)=\sinh \bar{t},\cosh \bar{t}$ in which
case the LRS metric becomes homogeneous. This result invalidates the
statement \cite{Kramer} that no $G_5$ exists on $V_4$ for the spacetime (\ref
{sx1.3}).

{\bf Case B}

From (\ref{sx4.1}) we obtain the condition $LL_{,\bar{t}\bar{t}}-(L_{,\bar{t}%
})^{2}=0$ which has the solutions:

\[
\begin{array}{c}
L(\bar{t})=c\qquad \qquad A(\bar{t})=cB(\bar{t}) \\ 
\\ 
L(\bar{t})=ce^{\bar{t}/c}\qquad \qquad A(\bar{t})=ce^{\bar{t}/c}B(\bar{t}).
\end{array}
\]
For both solutions the metric (\ref{sx4.2}) corresponds to the Minkowski
spacetime.

\underline{$L(\overline{t})=c$}

The 2-metric $ds_{L}^{2}$ becomes $ds_{L}^{2}=c^{2}e^{-2x}(-d\bar{t}%
^{2}+dx^{2})$ and by means of the transformation $\hat{t}=ce^{-x}\sinh 
\overline{t},$ $\hat{x}=ce^{-x}\cosh \overline{t}$ the reduced metric $%
ds_{2+2}^{2}$ takes its canonical form $-d\hat{t}^{2}+d\hat{x}%
^{2}+dy^{2}+dz^{2}$ whose conformal algebra is known \cite{Bruhat}. The CKVs
of the LRS metric can be written immediately in the original co-ordinates
and we do not refer them explicitly. All 11 CKVs are proper and there exists
an {\em inheriting proper} CKV, the ${\bf \xi =}\partial _{\bar{t}}$. This
CKV reduces to the HVF ${\bf \xi =}bt\partial _{t}$ with conformal factor $b$
provided that $A(t)=bt$ and $B(t)=\frac{b}{c}t$ ($b,c\neq 0$). This result
has been obtained previously by Sintes \cite{Sintes} however with the
unnecessary assumption of perfect fluid matter.

\underline{$L(\overline{t})=ce^{\bar{t}/c}$}

In this case the 2-metric $ds_{L}^{2}$ becomes $ds_{L}^{2}=c^{2}e^{2(\frac{%
\bar{t}}{c}-x)}(-d\bar{t}^{2}+dx^{2})$ and by means of the transformation:

\begin{eqnarray*}
\hat{t} &=&\frac{c^{2}}{2}\left[ \frac{1}{c+1}\exp \{\frac{c+1}{c}(\bar{t}%
-x)\}-\frac{1}{c-1}\exp \{\frac{c-1}{c}(\bar{t}+x)\}\right] \\
\hat{x} &=&\frac{c^{2}}{2}\left[ \frac{1}{c+1}\exp \{\frac{c+1}{c}(\bar{t}%
-x)\}+\frac{1}{c-1}\exp \{\frac{c-1}{c}(\bar{t}+x)\}\right]
\end{eqnarray*}
for $c\neq 1$ and:

\[
\hat{t}=\frac{1}{2}\left[ \frac{1}{2}\exp \{2(\bar{t}-x)\}+\bar{t}+x\right]
\qquad \hat{x}=\frac{1}{2}\left[ \frac{1}{2}\exp \{2(\bar{t}-x)\}-(\bar{t}%
+x)\right] 
\]
for $c=1$ the metric $ds_{2+2}^{2}$ takes the standard form $-d\hat{t}^{2}+d%
\hat{x}^{2}+dy^{2}+dz^{2}$. The comments on the CKVs of the previous case $L(%
\overline{t})=c$ apply. From the 11 proper CKVs the {\em proper }CKV ${\bf %
\xi =}c\partial _{\bar{t}}+\partial _{x}$ with conformal factor $\psi ({\bf X%
}_{1})=1+c(\ln B)_{,\bar{t}}$ is {\em inheriting. }Moreover only this CKV
can be reduced to HVF ${\bf \xi =}bt\partial _{t}+\partial _{x}$ with
conformal factor $b$, in which case the metric functions are given by $A(t)=%
\frac{b}{c}t$ and $B(t)=t^{1-\frac{1}{b}}$.

\section{The CKVs of the LRS Class I,III metrics}

\setcounter{equation}{0}

The LRS metrics of Class I and III defined in (\ref{sx1.2}) are different
from the ones of Class II considered previously because they are not
conformal to a 2+2 decomposable metric. Hence we have to develop a new
method and most helpful in this direction is the important Theorem of
Defrise-Carter. This Theorem demands that the conformal algebra of a metric
of Petrov type D is isomorphic to the Killing algebra of a conformally
related metric \cite{Defrise-Carter,Hall-Steele}. Because the LRS\ \label%
{second}Class I and III metrics are of Petrov type D the Theorem applies.
Due to the fact that the results for Class I ($\varepsilon =1$) spacetimes
follow from those of Class III spacetimes ($\varepsilon =-1$) with the
interchange of the co-ordinates $t,x$, we need to consider Class III only.

The KVs which span the $G_{4}$ are:

\[
{\bf K}_1=\partial _x,\qquad {\bf K}_2=\partial _z 
\]
\begin{equation}
{\bf K}_3=f(y,k)\cos z\partial _x+\sin z\partial _y+\left[ \ln \Sigma
(y,k)\right] _{,y}\cos z\partial _z  \label{sx6.1}
\end{equation}
\[
{\bf K}_4=f(y,k)\sin z\partial _x-\cos z\partial _y+\left[ \ln \Sigma
(y,k)_{,y}\right] \sin z\partial _z 
\]
where: 
\begin{equation}
f(y,k)=\Lambda (y,k)\left[ \ln \frac{\Lambda (y,k)}{\Sigma (y,k)}\right]
_{,y}  \label{sx6.2}
\end{equation}
They have the Lie brackets:

\begin{eqnarray}
\lbrack {\bf K}_{1},{\bf K}_{2}] &=&0,\qquad [{\bf K}_{1},{\bf K}%
_{3}]=0,\qquad [{\bf K}_{1},{\bf K}_{4}]=0,\qquad [{\bf K}_{2},{\bf K}_{3}]=-%
{\bf K}_{4}  \nonumber \\
&&  \label{sx6.3} \\
\lbrack {\bf K}_{2},{\bf K}_{4}] &=&{\bf K}_{3},\qquad [{\bf K}_{3},{\bf K}%
_{4}]=-k{\bf K}_{2}+2(1-k^{2}){\bf K}_{1}.  \nonumber
\end{eqnarray}
Let us assume that the metrics (\ref{sx1.2}) admit exactly one more CKV, the 
${\bf X}_{I}$ say, which together with the $G_{4}$, generates a conformal
group $G_{5}$. According to the theorem of Defrise-Carter (and the structure
of the isometry $G_{4}$) the metric (\ref{sx1.2}) must be conformally
related to another LRS metric, say $d\hat{s}^{2}$, of the form (\ref{sx1.2})
for which the conformal group $G_{5}$ is a group of isometries. Considering
the commutator of ${\bf X}_{I}$ with the KVs ${\bf K}_{1},{\bf K}_{2},{\bf K}%
_{3},{\bf K}_{4}$ and using Jacobi identities and Killing equations we
compute easily the form of ${\bf X}_{I}$ and the reduced homogeneous metric $%
d\hat{s}^{2}$. Finally returning to the original metric we obtain the
required CKV.

The assumption that the nonconformally flat LRS metrics (\ref{sx1.2}) admit
a conformal algebra $C_{6}$ or $G_{7}$ leads either to a symmetric spacetime%
\footnote{%
It is easy to show that the LRS space-times (\ref{sx1.2}) cannot be
symmetric ($R_{abcd;e}=0$) for any values of the metric functions $A(t)$ and 
$B(t)$.} or to a Petrov type N spacetime \cite{Hall-Steele}.

These results are stated in the following:

{\bf Proposition 2} {\em Nonconformally flat Ellis Class I,III LRS
spacetimes admit at most one proper and inheriting CKV given by: } 
\begin{equation}
{\bf X}_{I}=\partial _{\tilde{t}}+2ax\partial _{x}+ay\partial _{y}%
\mbox{
with }\psi ({\bf X}_{I})=a+(\ln B)_{,\tilde{t}}  \label{sx6.4}
\end{equation}
{\em in which case the metric functions are given by: }

\begin{equation}
\frac{A(t)}{B(t)}=c_{1}e^{-a\tilde{t}}\qquad h(t)=c_{2}B(t)e^{a\tilde{t}}
\label{sx6.6}
\end{equation}
\begin{equation}
\tilde{t}=\int \frac{dt}{h(t)}  \label{sx6.5}
\end{equation}
{\em where }$a=0${\em \ when }$k\neq 0${\em \ and }$c_{1},c_{2}${\em \ are
nonvanishing constants. Furthermore }$c_{1}\neq 1${\em \ for }$a=0${\em \ in
order to avoid the conformally flat case.}

The CKV (\ref{sx6.4}) reduces to a HVF with homothetic factor $b$ when $%
A(t)=d_{1}t^{\frac{b-2a}{b}}$, $B(t)=d_{2}t^{\frac{b-a}{b}}$ (found
previously in \cite{Sintes}) and to a KV when the metric functions $%
A(t)=d_{1}e^{-2at}$, $B(t)=d_{2}e^{-at}$.

We turn now to the conformally flat Class III metrics (\ref{sx1.2}). The
vanishing of the Weyl tensor is equivalent to the requirements $A(t)=B(t)$
and $k=-\varepsilon $. These conditions imply that the conformally flat
metrics (\ref{sx1.2}) are conformally related to the 1+3 decomposable
spacetimes: 
\begin{equation}
ds^{2}=B^{2}(t)\left\{ \varepsilon \left[ d\bar{t}^{2}-dz^{2}-2\Lambda
(y,-\varepsilon )dxdz-dx^{2}\right] +dy^{2}\right\}  \label{sx6.8}
\end{equation}
where $\bar{t}=\int \frac{dt}{B(t)}$ and whose 3-spaces: 
\begin{equation}
ds_{3}^{2}=\varepsilon \left[ -dz^{2}-dx^{2}-2\Lambda (y,-\varepsilon
)dxdz\right] +dy^{2}  \label{sx6.9}
\end{equation}
are spaces of constant curvature $R_{3}=-\frac{2}{3}\varepsilon .$ This
implies that the metrics (\ref{sx6.8}) are conformally related either to the
Einstein spacetime ($\varepsilon =-1$) or to RT spacetime ($\varepsilon =1$) 
\cite{RT}.

Using standard methods the conformal algebra of the 3-spaces (\ref{sx6.9})\
is computed easily from the conformal equations ${\cal L}_{{\bf K}%
}ds_{3}^{2}=2\phi ({\bf K})ds_{3}^{2}$ and it is listed in Table 5.
\pagebreak

{\bf Table 5. }The conformal algebra of the 3-metric (\ref{sx6.9}). The
index $\alpha =x,y,z$, $\varepsilon =\pm 1.$

\[
\begin{tabular}{|l|l|}
\hline
{\bf CKVs} & {\bf Conformal Factor }$\phi ({\bf K})=\frac \varepsilon
4\lambda $ \\ \hline
${\bf K}_1=\partial _x$ & \multicolumn{1}{|c|}{$0$} \\ \hline
${\bf K}_2=\partial _z$ & \multicolumn{1}{|c|}{$0$} \\ \hline
${\bf K}_3=f(y,-\varepsilon )\cos z\partial _x+\sin z\partial _y+\left[ \ln
\Sigma (y,-\varepsilon )\right] _{,y}\cos z\partial _z$ & 
\multicolumn{1}{|c|}{$0$} \\ \hline
${\bf K}_4=f(y,-\varepsilon )\sin z\partial _x-\cos z\partial _y+\left[ \ln
\Sigma (y,-\varepsilon )\right] _{,y}\sin z\partial _z$ & 
\multicolumn{1}{|c|}{$0$} \\ \hline
${\bf K}_5=\left[ \ln \Sigma (y,-\varepsilon )\right] _{,y}\cos x\partial
_x+\sin x\partial _y+f(y,-\varepsilon )\cos x\partial _z$ & 
\multicolumn{1}{|c|}{$0$} \\ \hline
${\bf K}_6=\left[ \ln \Sigma (y,-\varepsilon )\right] _{,y}\sin x\partial
_x-\cos x\partial _y+f(y,-\varepsilon )\sin x\partial _z$ & 
\multicolumn{1}{|c|}{$0$} \\ \hline
$K_{(7)\alpha }=\lambda _{1,\alpha }$ & $\lambda _1=\left[ \Lambda
(y,-\varepsilon )+1\right] ^{1/2}\sin (\frac x2+\frac z2)$ \\ \hline
$K_{(8)\alpha }=\lambda _{2,\alpha }$ & $\lambda _2=\left[ \Lambda
(y,-\varepsilon )+1\right] ^{1/2}\cos (\frac x2+\frac z2)$ \\ \hline
$K_{(9)\alpha }=\lambda _{3,\alpha }$ & $\lambda _3=\left[ 1-\Lambda
(y,-\varepsilon )\right] ^{1/2}\sin (\frac x2-\frac z2)$ \\ \hline
$K_{(10)\alpha }=\lambda _{4,\alpha }$ & $\lambda _4=\left[ 1-\Lambda
(y,-\varepsilon )\right] ^{1/2}\cos (\frac x2-\frac z2)$ \\ \hline
\end{tabular}
\]

\medskip

From the results of Table 5 we compute in a straightforward manner the CKVs
of the LRS spacetimes (\ref{sx6.8}) using the method developed in \cite
{Tsa-Nik-Apos}. The result is that the vectors ${\bf K}_{1},...,{\bf K}_{6}$
are KVs, the vector $\partial _{\bar{t}}$ is a proper CKV and the remaining
8 {\em proper} CKVs are:

\begin{eqnarray}
X_{(n)a} &=&-2\cos (\frac{\bar{t}}{2})\lambda _{(n)}\delta _{a}^{\bar{t}%
}-4\varepsilon \sin (\frac{\bar{t}}{2})\lambda _{(n),\alpha }\delta
_{a}^{\alpha }  \nonumber \\
&&  \label{sx6.10} \\
\psi ({\bf X}_{(n)}) &=&\lambda _{(n)}\left[ \sin (\frac{\bar{t}}{2})-2\cos (%
\frac{\bar{t}}{2})(\ln B)_{,\bar{t}}\right]  \nonumber
\end{eqnarray}

\begin{eqnarray}
X_{(n+4)a} &=&2\sin (\frac{\bar{t}}{2})\lambda _{(n)}\delta _{a}^{\bar{t}%
}+4\varepsilon \cos (\frac{\bar{t}}{2})\lambda _{(n),\alpha }\delta
_{a}^{\alpha }  \nonumber \\
&&  \label{sx6.11} \\
\psi ({\bf X}_{(n+4)}) &=&\lambda _{(n)}\left[ \cos (\frac{\bar{t}}{2}%
)+2\sin (\frac{\bar{t}}{2})(\ln B)_{,\bar{t}}\right]  \nonumber
\end{eqnarray}
where $n=1,2,3,4$ and $a=\bar{t},x,y,z$.

\section{LRS perfect fluid spacetimes admitting proper CKVs}

\setcounter{equation}{0}

The results of the previous sections are geometrical and apply to any type
of matter. However most existing (hypersurface homogeneous) LRS\ solutions
concern perfect fluids \cite{Kramer} and even recently new efforts are made
in the determination of new LRS\ perfect fluid solutions \cite
{Elst-EllisI,Marklund1,Marklund2,Kitamura1}. Hence it is of interest to
utilize the geometrical results of the previous sections and determine
explicitly all perfect fluid and nonconformally flat LRS spacetimes which
admit {\em proper CKVs}. The conformally flat cases are ignored because they
lead either to generalized Friedmann type metrics or to generalized interior
Schwarzschild type metrics which are well-known \cite{Kramer}.

The equivalence of the geometric results for $\varepsilon =+1$ and $%
\varepsilon =-1$ under the interchange of the coordinates $x\leftrightarrow
t $ is not transferred over to physics due to the timelike character of the
4-velocity. Thus we have to distinguish between the nonstatic case ($%
\varepsilon =+1$) and the static case ($\varepsilon =-1$).

The outline of the method of work is as follows. For each LRS metric
admitting CKVs we compute the energy momentum tensor $T_{ab}$. Using the
standard decomposition of the energy momentum tensor we determine the
dynamical quantities $\mu $, $p$, $q_{a}$ and $\pi _{ab}$ via the relations:

\begin{equation}
\mu =T_{ab}u^{a}u^{b},\mbox{ }p=\frac{1}{3}T_{ab}h^{ab},\mbox{ }%
q_{a}=-h_{a}^{c}T_{cd}u^{d},\pi _{ab}=h_{a}^{c}h_{b}^{d}T_{cd}-\frac{1}{3}%
(h^{cd}T_{cd})h_{ab},  \label{sx7.1}
\end{equation}
where $h_{ab}=g_{ab}+u_{a}u_{b}$ is the projection tensor associated with
the fluid four velocity $u_{a}$ ($u^{a}u_{a}=-1$). Assuming a perfect fluid
spacetime (i.e. $\pi _{ab}=0$ and $q_{a}=0$) we obtain a differential
equation which fixes the metric function $B(t)$ and consequently the exact
form of the dynamical variables. We demand that the weak and dominant energy
conditions $\mu >0$, $\mu +p>0$ and $\mu -p>0$ hold and select from the
resulting perfect fluid metrics those which are physically acceptable.

The form of the energy momentum tensor implies that for a perfect fluid
interpretation, the fluid velocity {\em must} be orthogonal to the group
orbits (nontilted models) i.e. $u^{a}\propto \delta _{\tau }^{a}$ except in
the case of the nonconformally flat ($\Leftrightarrow $ $A(t)\neq cB(t))$
metrics (\ref{sx1.3}) where a tilted 4-velocity is required. Hence we have
that:

{\bf Proposition 3 }{\em All hypersurface homogeneous LRS metrics which
admit a proper CKV have a perfect fluid interpretation for nontilted
observers, except the nonconformally flat metrics (\ref{sx1.3}) which admit
a perfect fluid interpretation only for tilted observers.}

In Tables 6,7 we collect the results of the calculations for the nontilted,
static and the nonstatic perfect fluid LRS metrics which admit a proper CKV.
\pagebreak

{\bf Table 6. }List of all nonconformally flat, nontilted and {\em static}
perfect fluid LRS spacetimes admitting {\em proper} conformal symmetries and
satisfying the weak and the dominant energy conditions. For the Ellis class
II the metric is $ds^{2}=B^{2}(\bar{x})\left[ -C^{2}(\bar{x})dt^{2}+d\bar{x}%
^{2}+dy^{2}+\Sigma ^{2}(y,k)dz^{2}\right] $ where $B(\bar{x}),C(\bar{x})$
are smooth functions given in the third and fourth column of the Table and
we have used the transformation $d\bar{x}=\frac{dx}{B(x)}$. For the Ellis
class I the metric is $B^{2}(\bar{x})\left[ c_{2}^{2}e^{2a\tilde{x}}d\tilde{x%
}^{2}-c_{1}^{2}e^{-2a\tilde{x}}\left( dt+\Lambda ^{2}(y,k)dz\right)
^{2}+\left( dy^{2}+\Sigma ^{2}(y,k)dz^{2}\right) \right] $ and we have used
the transformation $dx=c_{2}B(\bar{x})e^{a\bar{x}}d\tilde{x}$ where $%
c_{1},c_{2}$ are constants of integration. Furthermore $\varepsilon _{1}=\pm
1$ and $h(\bar{x},c)\equiv \sinh ^{-1}c\bar{x}$ or $\cosh ^{-1}c\bar{x}$.

\begin{tabular}{|l|l|c|c|c|l|l|}
\hline
{\bf 
\begin{tabular}{l}
Ellis \\ 
class
\end{tabular}
} & {\bf Case} & ${\bf k}$ & $B(\bar{x})$ & $C(\bar{x})$ & {\bf Constants} & 
{\bf Restrictions} \\ \hline
$II$ & A$_{1}$ & $0$ & $\frac{1}{\beta \bar{x}^{\frac{D+1}{2}}+\bar{x}^{-%
\frac{D-1}{2}}}$ & $\bar{x}^{\frac{\alpha _{1}-1}{\alpha _{1}}}$ & 
\multicolumn{1}{|c|}{$D=\frac{\sqrt{1+(\alpha _{1}-1)^{2}}}{\alpha _{1}}$} & 
$\alpha _{1}<1$ \\ \hline
$II$ & A$_{2}$ & $1$ & $\cosh ^{-1}\frac{\bar{x}}{\sqrt{2}}$ & $1$ & $-$ & 
none \\ \hline
$II$ & A$_{3}$ & $1$ & $h(\bar{x},c)$ & $e^{\varepsilon _{1}\bar{x}/a}$ & $c=%
\sqrt{\frac{1+a^{2}}{2a^{2}}}$ & $a\neq 1,a^{2}>1$ \\ \hline
$II$ & A$_{3}$ & $-1,0$ & $\cosh ^{-1}c\bar{x}$ & $e^{\varepsilon _{1}\bar{x}%
/a}$ & $c=\sqrt{\frac{1+ka^{2}}{2a^{2}}}$ & $a\neq 1,a^{2}<1$ \\ \hline
$II$ & A$_{4}$ & $1$ & $h(\bar{x},c)$ & $\cosh (\bar{x}/a)$ & $c=\sqrt{\frac{%
1+a^{2}}{2a^{2}}}$ & $a\neq 1$ \\ \hline
$II$ & A$_{4}$ & $-1,0$ & $\cosh ^{-1}c\bar{x}$ & $\cosh (\bar{x}/a)$ & $c=%
\sqrt{\frac{1+ka^{2}}{2a^{2}}}$ & $a^{2}<1/2$ \\ \hline
$II$ & A$_{5}$ & $1$ & $h(\bar{x},c)$ & $\sinh (\bar{x}/a)$ & $c=\sqrt{\frac{%
1+a^{2}}{2a^{2}}}$ & $a\neq 1,a^{2}>1$ \\ \hline
$II$ & A$_{6}$ & $1$ & $h(\bar{x},c)$ & $\cos (\bar{x}/a)$ & $c=\sqrt{\frac{%
1-a^{2}}{2a^{2}}}$ & $a\neq 1,a^{2}>1$ \\ \hline
$II$ & A$_{6}$ & $1$ & $\bar{x}^{-1}$ & $\cos \bar{x}$ & $-$ & none \\ \hline
$II$ & A$_{7}$ & $1$ & $h(\bar{x},c)$ & $\bar{x}$ & $c=\frac{\sqrt{2}}{2}$ & 
none \\ \hline
$I$ & $-$ & $0$ & $e^{-a\tilde{x}/2}\cosh (\frac{D\tilde{x}}{2})$ & $-$ & $D=%
\sqrt{5a^{2}+4c_{1}^{2}c_{2}^{2}}$ & none \\ \hline
$I$ & $-$ & $\pm 1$ & $\cosh (\frac{D\tilde{x}}{2})$ & $-$ & $D=c_{2}\sqrt{%
c_{1}^{2}+2k}$ & $c_{1}^{2}+2k>0,c_{1}\neq 1$ \\ \hline
$I$ & $-$ & $-1$ & $\cos (\frac{D\tilde{x}}{2})$ & $-$ & $D=c_{2}\sqrt{%
2-c_{1}^{2}}$ & $c_{1}^{2}<2,c_{1}\neq 1$ \\ \hline
\end{tabular}
\vspace{2cm}

{\bf Table 7. }List of all nonconformally flat, nontilted and {\em nonstatic}
perfect fluid LRS spacetimes admitting {\em proper} conformal symmetries and
satisfying the weak and the dominant energy conditions. The spacetime metric
is $ds^{2}=B^{2}(\tau )\left[ -d\tau ^{2}+C^{2}(\tau )dx^{2}+dy^{2}+\Sigma
^{2}(y,k)dz^{2}\right] $ and we have introduced the new time co-ordinate $%
d\tau =\frac{dt}{B(t)}$. Also $m$ is a nonvanishing natural number.

\begin{center}
\begin{tabular}{|l|l|c|c|c|l|l|}
\hline
{\bf Ellis class} & {\bf Case} & ${\bf k}$ & $B(\tau )$ & $C(\tau )$ & {\bf %
Constants} & {\bf Restrictions} \\ \hline
$II$ & A$_{4}$ & $-$ & $\sinh ^{c}\frac{\tau }{a}$ & $\cosh (\tau /a)$ & $c=%
\frac{ka^{2}-1}{2}$ & $a^{2}>1$ \\ \hline
$II$ & A$_{5}$ & $-$ & $\cosh ^{c}\frac{\tau }{a}$ & $\sinh (\tau /a)$ & $c=%
\frac{ka^{2}-1}{2}$ & $a^{2}>1$ \\ \hline
$II$ & A$_{6}$ & $1$ & $\sin ^{c}\frac{\tau }{a}$ & $\cos (\tau /a)$ & $c=-%
\frac{ka^{2}+1}{2}$ & $a^{2}=2m+1$ \\ \hline
$II$ & A$_{7}$ & $\pm 1$ & $c_{1}e^{\frac{k\tau ^{2}}{4}}$ & $\tau $ & $-$ & 
none \\ \hline
\end{tabular}
\bigskip
\end{center}

It is interesting to analyze qualitatively the basic physical properties of
spacetimes listed in Tables 6,7. We restrict our considerations to the
nonstatic LRS spacetimes which, in principle, can be used as cosmological
models and we examine whether these models inflate ($q<0$) and/or isotropise
($\sigma /\theta \rightarrow 0$ as $\tau \rightarrow \infty $) \cite
{Collins-Hawking,Wainright-Ellis} where $\theta =u_{;a}^{a}$ is the rate of
volume expansion, $\sigma $ is the shear and $q=\left( \frac{3}{\theta }%
\right) _{,\tau }-1$ is the deceleration parameter. For these spacetimes
(Table 7)\ we find that the only models which isotropise are those of the
case $A_{7}$ for both values $k=\pm 1$ ($k\neq 0$). The kinematical and the
dynamical parameters of these perfect fluid cosmological models are
collected in Table 8.\bigskip

{\bf Table\ 8. }

\begin{center}
\begin{tabular}{|c|c|c|c|c|c|}
\hline
$k$ & $\mu $ & $p$ & $\sigma $ & $\theta $ & $q$ \\ \hline
$1$ & $\frac{e^{-\tau ^{2}/2}(3\tau ^{2}+8)}{4c_{1}^{2}}$ & $-\frac{e^{-\tau
^{2}/2}(\tau ^{2}+8)}{4c_{1}^{2}}$ & $\frac{\sqrt{3}e^{-\tau ^{2}/4}}{%
3c_{1}^{2}\tau }$ & $\frac{e^{-\tau ^{2}/4}(3\tau ^{2}+2)}{2c_{1}^{2}\tau }$
& $3c_{1}^{2}e^{\tau ^{2}/4}\frac{(3\tau ^{4}-4\tau ^{2}+4)}{(3\tau
^{2}+2)^{2}}-1$ \\ \hline
$-1$ & $\frac{e^{\tau ^{2}/2}(3\tau ^{2}-8)}{4c_{1}^{2}}$ & $-\frac{e^{\tau
^{2}/2}(\tau ^{2}-8)}{4c_{1}^{2}}$ & $\frac{\sqrt{3}e^{\tau ^{2}/4}}{%
3c_{1}^{2}\tau }$ & $\frac{e^{\tau ^{2}/4}(-3\tau ^{2}+2)}{2c_{1}^{2}\tau }$
& $3c_{1}^{2}e^{-\tau ^{2}/4}\frac{(3\tau ^{4}+4\tau ^{2}+4)}{(3\tau
^{2}-2)^{2}}-1$ \\ \hline
\end{tabular}
\bigskip
\end{center}

From the results listed in Table 8 we draw the following conclusions:

a. For the nonstatic spherically symmetric ($k=1$) LRS spacetimes the energy
density never vanishes. For $\mid c_{1}\mid <2$ and $\tau \epsilon (0,\tau
_{0})$, where $\tau _{0}$ is a constant, $q<0$ and the model inflates.

b. For the nonstatic LRS spacetimes of case $A_{7}$ with $k=-1$ the energy
density vanishes at the value $\tau _{1}=\sqrt{8/3}.$ Hence there is always
a cosmological singularity of Kasner type at a finite time in the past.
Furthermore these indefinitely expanding models ($\theta \rightarrow \infty $%
) inflate because the deceleration parameter $q$ is negative for $\tau >\tau
_{0}$.

Furthermore from Table\ 8 we observe that the fluid has a nonlinear
barotropic equation of state.

{\bf Proposition 4 }{\em Nonstatic LRS perfect fluid spacetimes with linear
equation of state do not admit proper CKVs}.

It should be noted that for both models the ratio $p/\mu \rightarrow -1/3.$
Thus, asymptotically, the equation of state is linear i.e. $p=(\gamma -1)\mu 
$ with $\gamma =2/3$.

\section{Conclusions}

\setcounter{equation}{0}

We have determined all hypersurface homogeneous, locally rotationally
symmetric spacetimes which admit CKVs (including KVs and HVFs) together with
the explicit expression of these vectors and their conformal algebras. These
results take previous studies of the same topic one step further and have
been applied in two directions.

(a) To compute and classify the conformal algebra of {\em all static} LRS
spacetimes (including the interesting case of static spherically symmetric
spacetimes) in a simple, new and geometrical way

(b) To determine all perfect fluid spacetimes (that is, perfect fluid models
which satisfy the weak and the dominant energy conditions) which admit
proper CKVs. In addition we have shown that there is a class of nonstatic
LRS spacetimes admitting conformal symmetries which isotropise at late times
and, in principle, they can be used as physically reasonable cosmological
models. These results confirm the fact that conformal symmetries may play an
important role in cosmology and they provide an additional motivation of
studying models admitting conformal symmetries.

The geometrical approach we used has made possible the determination of all
the inheriting CKVs of the LRS spacetimes independently of the matter
content. Concerning these vectors we note that in the Case A$_{1}$ of Ellis
Class II spacetimes (\ref{sx1.1}) the inheriting CKV ${\bf \xi }$ shares the
common property with the inheriting CKVs of the Robertson-Walker metric \cite
{Coley-Tupper4,Maa-Mah} that is, they reduce to a HVF and/or KVs. The
inheriting CKV ${\bf \xi }_{2}$ of cases A$_{2}$ and A$_{3}$ is known. The
former belongs to the Robertson-Walker case of inheriting CKVs found by
Coley and Tupper \cite{Coley-Tupper4}. For the special case $k=+1$ we
recover the inheriting CKVs and the associated spherically symmetric,
anisotropic in general, spacetimes determined by Coley and Tupper. More
precisely cases A$_{2}$ and A$_{3}$ of the present paper correspond to the
classes $A_{2}S$ and $A_{1}S$ of \cite{Coley-Tupper5} respectively.
Regarding the inheriting CKVs of the metrics (\ref{sx1.2}), (\ref{sx1.3}) we
have seen that the first admits only one inheriting CKV, the ${\bf X}_{I}$
and the second admits the inheriting proper CKV ${\bf X}_{1}$ provided that
the fluid velocity is nontilted. These results are new and can be used to
study the kinematical and the dynamical properties of the associated metrics.

{\bf Acknowledgments}

One of the authors (PSA) was partial supported by the Hellenic Fellowship
Foundation (I.K.Y.).

\end{document}